\documentstyle[prb,aps,twocolumn]{revtex}
\def\eq{\begin{equation}}
\def\be{\begin{equation}}
\def\ee{\end{equation}}
\def\eqa{\begin{eqnarray}}
\def\eea{\end{eqnarray}}
\def\ra{\rightarrow} 
\def\rxx{\rho_{xx}}
 
\def\sxx{\sigma_{xx}}
\def\sxy{\sigma_{xy}} 
\def\s{\sigma}
\def\p{\partial}
\def\psib{\bar{\psi}}
\def\bb{\bar{b}}
\def\rb{\bar{\rho}}     
\def\r{\rho}                              
\def\w{\omega}

\def\ep{\epsilon}
\def\bt{\beta}
\def\ap{\alpha}

\begin{document}
\draft
\flushbottom
\twocolumn[
\hsize\textwidth\columnwidth\hsize\csname @twocolumnfalse\endcsname

\title{Composite Fermion Hall Conductance at $\nu=1/2$}
\author{D-H Lee$^{(a)}$, Y. Krotov$^{(a)}$, J. Gan$^{(a)}$ and S.A. Kivelson$^{(a,b)}$}
\address{$^{(a)}$
Department of Physics, University of California at Berkeley, 
Berkeley, CA 94720}
\address{$^{(b)}$Dept. of Physics
University of California at Los Angeles
Los Angeles, CA 90095} 
\date{\today}
\maketitle
\tightenlines
\widetext
\advance\leftskip by 57pt
\advance\rightskip by 57pt

\begin{abstract}
We show that in the 
limit of vanishing bare electron effective mass, and in the presence of 
particle-hole symmetric disorder (which can be of vanishing
strength), the composite fermion Hall 
conductance is constrained to be $-{1\over 2}\frac{e^2}{h}$.  We discuss the
implications of this results for the existence and nature of a composite
Fermi liquid in the lowest Landau level. 
\end{abstract}

\vskip 1cm

]

\narrowtext
\tightenlines

The observation of a seemingly metallic DC magnetotransport\cite{jiang} and
the subsequent discovery of an acoustic wave anomaly\cite{willet} near 
$\nu=1/2$, opened a new chapter in the studies of quantum Hall effects. 
(Here $\nu\equiv \phi_0\rb/B$, where $\rb$ is the mean electron density, 
$\phi_0 =hc/e$, and $B$ is the externally applied magnetic field.) 
A {\it very} intriguing idea, the composite fermion theory, has been put forward to 
explain these phenomena.\cite{hlr,kz}
In this theory, each electron is represented as a 
composite-fermion\cite{jain} carrying two quanta of fictitious magnetic flux 
which pierce the physical plane in the direction opposite to that of the real 
magnetic flux. 
Formally, this transformation maps the problem of electrons in a
strong magnetic field onto a system of ``composite
fermions'' moving in the same external field while interacting with a 
fluctuating ``statistical'' gauge field governed by 
a Chern-Simons action.\cite{zhk} 

{\it In the absence of disorder}, the ground state electron density 
is uniform. Thus at the {\it mean-field} level, the averaged statistical 
magnetic field, 
$\bb=2\phi_0\rb=|B|$, cancels the 
external one, and the composite-fermions see no net field.
When one tries to improve upon the mean-field theory(MFT) by 
including the fluctuations of the statistical magnetic 
field, one encounters divergences.\cite{hlr} 
Attempts to sum these divergence have led to suggestive, 
but so far inconclusive results.\cite{stern,wilczek,marsden,ioffe,stamp} 
Despite this difficulty, it has been {\it conjectured} that the full 
effect of statistical gauge field fluctuations is to renormalize the 
parameters (perhaps in a singular way) of a zero field ``composite Fermi 
liquid''. 

At this point it is useful to differentiate two concepts. The 
first is the composite fermion approach, and the second is the composite 
Fermi liquid theory. The former is simply an exact reformulation of the 
original problem, but the latter is a conjecture about the final solution.
It is also worth pointing out that although the magnetic field is canceled 
out at the mean-field level in the composite fermion approach, there is no 
symmetry reason to expect $\sxy^f=0$ since the full composite fermion action 
lacks time reversal symmetry. It is {\it our understanding} however, that the 
composite Fermi liquid conjecture requires that $\sxy^f=0$.  
  
In any case, it has been argued that the transport properties  of the 
electrons near 
$\nu=1/2$ simply reflect the underlying Fermi liquid (or, possibly, 
the marginal 
Fermi liquid) behavior of the composite fermions in {\it zero} magnetic field. 
This intriguing picture acquired further support when
Fermi-surface-like features were observed in recent experiments.\cite{fsurf}

The principal purpose of the present paper is to reexamine
the Fermi liquid picture when there is a finite (but 
possibly arbitrarily small) 
amount of disorder. In the presence of disorder, the ground state electron 
density is no longer uniform. In the regions of high electron density, the 
statistical magnetic flux over-compensates the external one, and in the low 
density region it under-compensates. Thus from the view point of the composite 
fermions, the plane is divided into regions with net effective fields 
opposite to 
each other. Nominally, if the average field is zero, one would expect a 
vanishing Hall conductance for composite fermion (i.e. $\sxy^f=0$). This 
naive expectation is {\it incorrect} because of the {\it correlation} 
between the composite fermion density (which is equal to the electron density) 
and the effective magnetic field.
Thus even if there are as 
many regions with positive and negative net magnetic field, 
one might expect the composite 
fermion Hall conductance to be negative (i.e. $\sxy^f<0$). The
existence of this correlation between flux and charge
also raises questions concerning the validity of models of composite
fermion transport in
which this correlation is ignored.\cite{kz}

Here, 
we shall concentrate on a particular limit, (the limit where the bare electron  
effective mass $m_b$ is vanishingly small, and the disorder potential is 
particle-hole symmetric\cite{note}), where we will show that at $\nu=1/2$ the {\it electron} Hall conductivity is, 
\be
\sigma_{xy}= {1\over 2}\frac{e^2}{h}.
\label{half}
\ee
The {\it electron} 
resistivity tensor is related to the {\it composite fermion} resistivity
tensor by a  ``connection formula'' (which will be discussed below) 
\eqa
&&\rxx=\rxx^f\nonumber \\
&&\r_{yx}=2\frac{h}{e^2}+\r_{yx}^f;
\label{connection}
\eea
combining this expression with Eq. (\ref{half}),
we will show that, so long as  $\sigma_{xx}\ne 0$,
it follows that
\eq
\sxy^f=-{1\over 2}\frac{e^2}{h},
\label{sxyf}
\ee 
independent of the strength of the disorder or whether the temperature is zero
or finite!

Disorder (or some other interaction which breaks Galilean invariance)
is {\it essential} to establish the above constraint 
on $\sxy^f$. Galilean invariance requires the 
{\it electron} resistivity tensor to be
\eqa
&&\rxx=0\nonumber \\
&&\rho_{yx}=2\frac{h}{e^2}.
\label{connect}
\eea
The above result combined with the connection formula Eq. (\ref{connect}) 
implies that
\eq
\rxx^f=\rho_{yx}^f=0.
\ee
In this latter case the composite Fermion resistivity
tensor is non-invertable.  Therefore,
our conclusions concerning $\sigma_{xy}^f$
apply in the limit that the disorder tends to zero, but may not
apply in the absence of disorder.

The remainder of the paper is organized as follows:  In Section I we
derive Eq. (\ref{half}).  In Section II, we show how Eq. (\ref{sxyf})
follows from Eq. (\ref{half}).  In the remainder of the paper, we
attempt to understand the implications of this result on the
fundamental character of the physical state at $\nu=1/2$.
Section III contains some discussion
of the nature of the ground-state in the presence of disorder in the
$m_b \rightarrow 0$ limit.  In Section IV, we examine the problem of
computing $\sigma_{xy}^f$ in the absence of disorder, but including
the perturbative effects of fluctuations about the mean-field state.
We find that, at least to lowest order, the mean field result
$\sigma_{xy}^f=0$ is unchanged.  We also discuss our reasons for believing
that, even though the present results are derived in a way that depends
critically on the existence of a disorder potential, they raise
important questions concerning the nature of the ground-state in the
lowest Landau level at $\nu=1/2$, even in the absence of disorder. 
Section V is a discussion of some other perturbative results in the absence
of disorder;  formally this section is a digression from the main thrust
of the paper, except in that it sheds some light on the nature of the
composite Fermion ground-state at $\nu=1/2$.  Section VI contains a 
discussion of results, some speculations concerning their possible
implications, and a discussion of their possible relevance to
experiment.  For the remainder of the paper,
we adopt units in which $e/c=k_B=\hbar=1$.

\section{The Hall conductance at $\nu=1/2$ in the limit of vanishing band mass}

\subsection{Intuitive discussion:  particle-hole symmetry in the lowest 
Landau level}

In the limit of small $m_b$, or equivalently when the cyclotron frequency, 
$\omega_c =B/m_b$ is the largest energy in the problem, we expect that
the low-lying eigenstates for $\nu \le 1$ can be constructed out of states
lying entirely in 
the lowest Landau level plus perturbative effects of Landau level
mixing. It is easy to see\cite{girvin} 
that even in the presence of electron-electron interactions and
particle-hole symmetric disorder,\cite{note} 
the Hamiltonian projected onto the lowest 
Landau-level is particle-hole symmetric.  This is roughly, but
not quite, adequate for our
present purposes.  What we seek to investigate is the nature of
this symmetry for the full problem,
in the physically meaningful limit $m_b \rightarrow 0$;  intuitively,
this limit is 
related to Landau-level projection, but there are effects of
Landau-level mixing which survive in this limit,\cite{sk}
especially when the current operator is involved.

None-the-less, we will start our discussion by assuming that particle-hole
symmetry is an exact low energy 
symmetry, and
discuss its consequences.  (In the following subsections we will
demonstrate that, subject to some reasonable assumptions, the inferences
we have made can be substantiated.)

Since the ground 
state at $\nu=1$ is unique,
it can play the role of a reference vacuum equally well as the state with
no electrons.  What this means is that a system with electron concentration
$\nu<1$ can be viewed, equivalently, as a system of holes with concentration
$1-\nu$.  The corresponding conductivity tensor as a function of filling 
factor can be expressed in either electron or hole language as
\be
\sigma(\nu)=\sigma(1) + \sigma^{h}(1-\nu)
\ee
where $\sigma^{h}(1-\nu)$ is the hole conductivity tensor at hole concentration
$1-\nu$ (electron concentration $\nu$).  Particle hole symmetry,
in turn, implies that
\be
\sigma_{xx}(\nu)=\sigma^{h}_{xx}(\nu)
\ee
and
\be
\sigma_{xy}(\nu)=-\sigma^{h}_{xy}(\nu).
\ee
>From these equations, we can exactly relate the Hall conductivity at $\nu=
1/2$ to the Hall conductivity at $\nu=1$:
\be
\sigma_{xy}(\nu=1/2)=(1/2)\sigma_{xy}(\nu=1).
\label{sxy1}
\ee
Since $\sigma_{xy}(\nu=1)=\frac{e^2}{2\pi}$, Eq. (\ref{sxy1}) implies Eq. (\ref{half}).

Equation (\ref{sxy1}) is a strong result, and it applies not only to the
D.C conductivity, but to finite frequency, $\omega$, finite wave number,
$\vec k$, and finite temperature, $T$, to the extent that none of these
are large enough to imply substantial Landau-level mixing and hence
breaking of particle-hole symmetry;  indeed, we expect corrections due to
finite temperature and finite frequency to vanish in the $\omega_c\rightarrow
\infty$ limit, and finite $k$ corrections to go like $(kl)^2$, where
$l=\sqrt{B}$ is the magnetic length.

\subsection{Particle-hole symmetry at zero temperature}

In this subsection, we show that at $T=0$ and in the limit $m_b \rightarrow 0$,
in the presence of particle-hole symmetric disorder\cite{note}
and under the assumptions that: 
i) as a function of $m_b$ there are no non-analytic 
contributions to the Hall-conductivity which survive
in the  
$m_b\rightarrow 0$ limit, and ii) there is no spontaneous particle-hole 
symmetry breaking, it follows that 
the electronic  
Hall conductivity is given by Eq. (\ref{half}).

The general expression for the Hall conductivity is given by the Kubo formula:
\eqa
&&\sigma_{xy}(\omega)
= \frac {A}{\omega } \int dt e^{i \omega t}
\theta (t) <<g[U]|[J_{x}(t),J_{y}(0)]|g[U]>>_{dis}\nonumber \\
&&
\label{kubo}
\eea
where $A$ is the total area, $\theta(t)$ is the Heavyside function,
and $J_{\alpha}$, the averaged current density, 
is given by
\begin{equation}
J_{\alpha}= \frac{e}{A}\int d^{2}r \frac{1}{2m_b}[\Psi^{\dag}(r)
(\frac{\partial_{\ap}}{i}-A_{\ap})\Psi(r)+h.c.].
\end{equation}
In the following we shall consider $\w<<\w_c$.
In Eq.(\ref{kubo}) 
$\mid g[U]>$ is the ground state for a given external potential $U(r)$, and 
$<...>_{dis}$ denotes the disorder average. (To simplify 
notation, we shall henceforth leave implicit the dependence of eigenstates 
on $U(r)$.)
Next we choose the eigenstates of the kinetic energy operator
as a basis and expand 
$\Psi(r) = \sum_{nk} \psi_{nk}(r) a_{nk}$
where $a_{nk}$  annihilates an electron in the state 
\begin{equation}
\psi_{nk} = \frac{1}{(Ll)^{\frac{1}{2}}} e^{iky} \chi_{n}(\frac{x-kl^{2}}{l}).
\end{equation}
Here we have chosen the gauge $\vec{A} = (0, Hx)$;
$L$ is the size of the system in the $y$ direction (Fig.\ \ref{hallbar}) and 
\begin{equation}
\chi_{n}(x) = (2^{n} n! \pi^{\frac{1}{2}} )^{- \frac{1}{2}} H_{n}(x) 
e^{-\frac{x^{2}}{2}},
\end{equation}
where $H_{n}(x)$ are the Hermit polynomials.
After some trivial algebra we obtain
\begin{equation}
J_{x} = \frac{e}{iAm_bl} \sum_{n,k} X_n [a^{\dag}_{nk} a_{n+1k}- a^{\dag}_{n+1k} a_{nk}],
\label{jx}
\end{equation}
where $ X_n = \int dx \chi_{n}(x) \partial_{x} \chi_{n+1}(x)$. Similarly,
\begin{equation}
J_{y} = - \frac{e}{Am_bl} \sum_{n,k} Y_n[ a^{\dag}_{nk} a_{n+1k}+ a^{\dag}_{n+1k} a_{nk}],
\label{jy}
\end{equation}
where $ Y_n = \int dx x\chi_{n}(x) \chi_{n+1}(x)$.  Note that $\vec{J}$ is
purely off-diagonal in the Landau-level index, but has non-zero matrix
elements only between neighboring Landau-levels.  In particular, for
our purposes, we need to know only the matrix elements, $X_0=Y_0=1/\sqrt{2}$.

Now let us consider the correlation function
\eqa
I([U];t)&=& <g|[J_{x}(t), J_{y}(0)]|g>\nonumber \\ &=&\sum_{\alpha}\{e^{-i(E_{\alpha} - E_{g})t}<g| J_{x}(0) |\alpha><\alpha|J_{y}(0)|g>\nonumber \\
&-& c.c \}
\label{corr}
\eea
where $\mid\ap>$ are the
true many-body eigenstates of the system in the presence of external disorder 
potential $U(r)$. To proceed, let us perform a canonical transformation
\eq
|\psi'>=e^{iT}|\psi>
\ee
so that the transformed Hamiltonian 
\eq
H'=e^{iT}He^{-iT}=H_{-1}+H_0+...
\label{hp}
\ee
has no matrix element connecting pure LLL states with those with a higher 
Landau level component.\cite{sk,sd}
The hermitian operator $T$ can be constructed as a series in
$m_b$ (which actually is an expansion in powers of the ratio of the Landau level mixing matrix element to $\omega_c$\cite{skk}), as follow
\eq
T=\sum_{k=1}^{\infty}(m_b)^kT_k.
\ee
Thus in Eq.(\ref{hp})   $H_k={\cal O}(m_b^k)$.
The transformed current operator has the expansion
\eq
\vec{J}'=e^{iT}\vec{J}e^{-iT}=\vec{J}_{-1}+\vec{J}_{0}+...,
\label{current}
\ee
where $\vec{J}_{-1}=\vec{J}$.

After the transformation, the eigenstates separate into two groups: 
one group $\{|\ap_l>\}$ lies entirely in the LLL, and the other 
$\{|\ap_h>\}$ contain higher Landau level components. (By assumption, 
$|g>\in\{|\ap_l>\}$.)
By construction, the lowest group ${|\ap_l>}$ are eigenstates of 
the projected Hamiltonian
$H_L=P_LH'P_L$, where $P_L$ is the operator that projects onto the subspace
spanned by states in the LLL. 
To ${\cal O}(m_b^0)$ 
\eqa
H_L&=&\mu\int d^2r\rho_L(r)+{1\over 2}
\int d^2rd^2r'
v(r-r'){:\rho_L(r)\rho_L(r'):}\nonumber \\
&+&\int d^2r U(r)\rho_L(r).
\label{h}
\eea
Here $\rho_L(r)=\psi_L^+(r)\psi_L(r)$, with $\psi_L(r)\equiv\sum_k
\psi_{0k}(r)a_{0k}$, and $U$ and $v$ are the disorder and interaction 
potential respectively.

Consider first  the contribution to Eq.(\ref{corr}) due to 
due to inter-Landau-level excitations: 
\begin{eqnarray}
I_{1}([U];t)&=& \sum_{\alpha_{h}} \{ e^{-i(E_{\alpha_h} - E_{g})t}
<g|J'_{x}(0) |\alpha_{h}> <\alpha_{h}|J'_{y}(0)|g> \nonumber\\
&-&c.c\}.
\label{ii1}
\end{eqnarray}
Since $|g>$ lies entirely in  
the lowest Landau level,
\eqa
&&<\alpha_{h}|J'_x|g>=\frac{eX_0}{iAm_bl}  
<\alpha_{h}|K^+|g>+{\cal O}(m_b^0)\nonumber \\
&&<\alpha_{h}|J'_y|g>= -\frac{eY_0}{Am_bl}
<\alpha_{h}|K^+|g>+{\cal O}(m_b^0),
\label{jjx}
\eea
where
\eq
K^+\equiv\sum_{k} a^{\dag}_{1k} a_{0k}.
\ee
Thus the corresponding contribution to $\sigma_{xy}$ is
\eqa
&&\sxy^{(1)}([U];\w)=\frac{A}{\w}\int dt\theta(t)e^{i\w t}I_1(t)
=\frac{e^2}{A(m_bl)^2}\times\nonumber \\
&&\sum_{\alpha_{h}}\frac{<g|K
|\alpha_{h}><\alpha_{h}|K^+|g>}{(E_{\ap_h}-E_g)^2-\w^2} + {\cal O}(m_b).
\label{s1}
\eea
Let us write 
\eq
E_{\ap}-E_g\equiv\w_c+\Delta_{\ap};
\label{gap}
\ee
to lowest order in $m_b$, we can approximate
$\Delta_{\ap}$ by 0
in Eq.(\ref{s1}). 
Thus, the leading order contribution to $\sigma_{xy}^{(1)}$ 
in the $m_b\rightarrow
0$ limit is
\eq
\sxy^{(1)}=\frac{e^2}{AB}<g|\sum_ka^+_{0k}a_{0k}|g>,
\label{iii1}
\ee
where we have used the fact that $(m_b l \omega_c)^2=B$.

Next we look at the contribution to $\sigma_{xy}$
due to intra-Landau-levels excitations.
\eq
\sxy^{(2)}([U];\w)=\frac{A}{\w}\int dt\theta(t)e^{i\w t}I_2(t),
\ee
where
\eqa
I_{2}([U];t)&=& \sum_{\alpha_{l}} \{<g| J'_{x}(t) |\alpha_{l}> 
<\alpha_{l}|J'_{y}(0)|g>\nonumber \\ 
&-& <g| J'_{y}(0) |\alpha_{l}> <\alpha_{l}|J'_{x}(t)|g>.
\label{iiii2}
\eea
To ${\cal O}(m_b^0)$, we can replace $\vec{J}'$ in Eq.(\ref{iiii2}) by
\eq
\vec{J}''\equiv P_L(\vec{J}_{-1} + \vec{J}_{0})P_L = 
P_L( \vec{J}_{0})P_L.
\ee
Thus,
\eq
I_2[U]=<g|[J_{x}''(t), J_{y}''(0)]|g> +{\cal O}(m_b),
\label{i2}
\ee
and
$J_{\ap}'' ={1\over A}\int d^2r j_{\ap}''(r)$ where\cite{sd}
\eqa
j_{\ap}''(r)&=& P_L j_{\ap}(r) \frac{1}{ \hbar \omega_{c} \hat{N}/2 - 
\hat{H}_{K} } (1-P_L)VP_L\nonumber \\
 &+& P_LV(1-P_L) \frac{1}{ \hbar \omega_{c} \hat{N}/2 - \hat{H}_{K}} 
j_{\ap}(r) P_L.
\label{jp}
\eea
Here $\hat{N}$ and $\hat{H}_{K}$ are the particle number operator and 
kinetic energy operator respectively, and
$V$ is the sum of the potential (disorder)
and the two-body interaction part of the Hamiltonian. The time dependent 
operator
$J_{x}''(t)$ is related to $J_{x}''(0)$ via
\eq
J_{x}''(t)=e^{itH_L}J_{x}''(0)e^{-itH_L},
\ee
Given Eq.(\ref{jp}) we perform the integration over space\cite{note2} 
and obtain: 
\eqa
J_{x}'' &=& \frac{e}{A} \int d^{2}r \{\rho_{L}(r) \partial_{y} U(r) + 
\int d^{2}r_{1} \rho_{L}(r) \rho_{L}(r_{1})\times\nonumber \\
&&\partial_{y} v(r-r_1)+\int d^{2}r_{1} \rho_{L}(r_{1}) \rho_{L}(r) 
\partial_{y} v(r_{1}-r) \},
\label{jpx}
\eea
and
\begin{eqnarray}
J_{y}'' &=& -\frac{e}{A} \int d^{2}r \{\rho_{L}(r) \partial_{x} U(r) - \int d^{2}r_{1} \rho_{L}(r) \rho_{L}(r_{1})\times\nonumber \\
&&\partial_{x} v(r-r_1)
-\int d^{2}r_{1} \rho_{L}(r_{1}) \rho_{L}(r) \partial_{x} v(r_{1}-r) \}.
\label{jpy}
\end{eqnarray}

At $\nu=1/2$, and when $\int d^2r U(r)=0$, the value of $\mu$ is such that 
$H_L[U]\ra H_L[-U]$ under the {\it LLL} p-h transformation,
\eq
\psi_L(r)\ra\psi^+_L(r).
\label{ph}
\ee
Eq.(\ref{ph}) amounts to the change
\begin{equation}
a_{0k} \rightarrow a^{\dag}_{0k},
\end{equation}
and complex conjugation of the basis wavefunction.
Under this transformation (since complex conjugation is equivalent to the
transformation $y\rightarrow -y$)
\begin{eqnarray}
J_{x}''(U) & \rightarrow & -J_{x}''(-U) \nonumber\\
J_{y}''(U) & \rightarrow & J_{y}''(-U).
\end{eqnarray}
{\it If}
\eq
\mid g_p[U]>\ra \mid g_P[-U]>
\label{break}
\ee
({\it i.e.} if there is no spontaneous particle-hole symmetry breaking),  
then (since $\sum_k [1]=AB/\phi_0=AB/2\pi$)
\eqa
&&\sxy^{(1)}[U]=\frac{e^2}{2\pi}-\sxy^{(1)}[-U]\nonumber \\
&&\sxy^{(2)}[U]= -\sxy^{(2)}[-U].  
\label{i1m}
\eea
Therefore
\eqa
&&\sxy\equiv<\sxy[U]>_{dis}=
\int D[U]P[U](\sxy^{(1)}[U]+\sxy^{(2)}[U])\nonumber \\
&&= {{e^2}\over {2\pi}}-\int D[U]P[U]\{
\sxy^{(1)}[-U]+\sxy^{(2)}[-U]\} \nonumber \\
&&={{e^2}\over {2\pi}}-\int D[U]P[-U]\sxy[U]\nonumber \\
&&={{e^2}\over {2\pi}}-\int D[U]P[U] \sxy[U]\nonumber \\
&&={{e^2}\over {2\pi}}-<\sxy[U]>_{dis}\nonumber \\
&&\equiv{{e^2}\over {2\pi}}-\sxy.
\label{i2m}
\eea
In the above equation we have used the fact that the disorder is particle-hole symmetric i.e.
\eq
P[U]=P[-U].
\ee
After restoring $\hbar$ Eq.(\ref{i2m}) is equivalent to Eq.(\ref{half}).

\subsection{Particle-hole symmetry at non-zero temperature}

The derivation presented above 
can be generalized to finite temperature.
In that case $<g|[J_x(t),J_y(0)]|g>$ in Eqs.(\ref{kubo}) and (\ref{corr}) 
is replaced by the thermal average, i.e.
\eq
<g|[J_x(t),J_y(0)]|g>\ra \frac{Tr\{e^{-\bt H}[J_x(t),J_y(0)]\}}{Tr\{e^{-\bt H}\}}.
\label{T}
\ee
By making  the same assumptions as in the above, we see that in the
$m_b \rightarrow 0$ limit,
we can evaluate the trace over states in the LLL. Thus Eq.(\ref{ii1})
is replace by
\begin{eqnarray}
&&I_{1}[U] =\frac{1}{Z_l}\sum_{\ap_l} e^{-\bt E_{\ap_l}}\times\nonumber \\
&&\sum_{\alpha_{h}}\{e^{-i(E_{\alpha_h} - E_{\ap_l})t}
<\ap_l | J'_{x}(0) |\alpha_{h}> <\alpha_{h}|J'_{y}(0)| \ap_l > \nonumber\\
&-&c.c\}.
\label{ii1p}
\end{eqnarray}
where 
\eq
Z_l\equiv \sum_{\ap_l} e^{-\bt E_{\ap_l}}.
\ee
To the lowest order in $m_b$, we again replace $\vec{J}'$ by $\vec{J}$
in Eq. (\ref{ii1p}).
Again, as we did in Eq. (\ref{gap}), we make the
replacement, valid to ${\cal O}(m_b^0)$, $(E_{\ap_h}-E_{\ap_l})\ra\w_c$. 
Then 
\begin{eqnarray}
&&\sxy^{(1)}=\frac{e^2}{AB}\frac{1}{Z_l}\sum_{\ap_l}e^{-\bt E_{\ap_l}}<\ap_l|\sum_ka^+_{0k}a_{0k}|\ap_l>.
\label{i1p}
\end{eqnarray}
Finally, Eq.(\ref{i2}) is replaced by
\eq
I_{2}[U]
=\frac{1}{Z_l}Tr'\{e^{-\bt H_L} [J_{x}''(t), J_{y}''(0)]\}.
\label{ii2}
\ee
Here $Tr'\{...\}\equiv\sum_{\ap_l}<\ap_l|...|\ap_l>$,
denotes the partial trace over the LLL eigenstates only.
At finite temperature the condition of no particle-hole symmetry breaking is generalized to the statement that we may use Eq.(\ref{T}) without including in $H$ an infinitesimal symmetry-breaking field.

\section{Implications for the composite Fermion conductivity}


An important ingredient of the composite fermion approach is the relation 
between the electron and composite fermion correlation functions.  It is the 
nature of the mapping that the density
of composite fermions equals that of the electrons, but the 
relation between current operators is more complicated.  To compute the 
electron current-current correlation function, we need to string together the composite fermion ``irreducible bubbles''\cite{klz} using the Chern-Simons 
bare gauge propagator. 
As shown in Refs.\cite{hlr},\cite{klz} and \cite{zhk}$^b$, 
this results in the relation
Eq.(\ref{connection}) 
between the resistivity tensor of electrons, $\rho_{\ap\bt}$, and 
that of the composite fermions, $\rho_{\ap\bt}^f$. 
Physically, this expresses the fact that associated with the
composite fermion current, there is a the statistical flux current, which 
produces a corresponding EMF proportional to the
statistical flux carried by 
each composite fermion times the electrical current.
When $\rxx^f\ne 0$, the resistivity tensor can be inverted
with the consequence that Eq.(\ref{connection}) is equivalent to 
\eqa
&&\sxy^f=\frac{e^2}{2\pi}\frac{\frac{e^2}{2\pi}\sxy-2(\sxx^2+\sxy^2)}{4\sxx^2+(\frac{e^2}{2\pi}-2\sxy)^2}\nonumber \\
&&\sxx^f=\frac{e^2}{2\pi}\frac{\frac{e^2}{2\pi}\sxx}{4\sxx^2+(\frac{e^2}{2\pi}-2\sxy)^2}.
\label{connection1}
\eea
In the above $\sxx$,and $\sxy$ are the {\it impurity averaged} conductivity tensor of the electrons. $\sxx^f,\sxy^f$ are the conductivity deduced from the {\it impurity averaged} bubble diagrams that are irreducible with respect to cutting
a statistical gauge propagator. Here we stress that
the latter is {\it not} necessarily equal to first taking the 
statistical-gauge-propagator-irreducible bubble in fixed disorder, 
and then averaging over the disorder realization. For example, the diagram
shown in 
Fig.\ref{irrbubble} belongs to the former, while not the latter.
By substituting $\sxy=\frac{e^2}{4\pi}$ into Eq.(\ref{connection1}), 
we obtain
\eqa
&&\sxx^f=(\frac{e^2}{4\pi})^2\frac{1}{\sxx}\nonumber \\
&&\sxy^f=-\frac{e^2}{4\pi}.
\label{cons}
\eea
The above 
is valid so long as $\rho_{xx} \ne 0$ and when
particle-hole symmetry is maintained, so
it applies with or without electron-electron interactions, for finite or 
infinite systems, and at zero or non-zero temperature.

\section{
What is the correct state in the limit of zero band mass in the presence of  
disorder?}

Now the remaining question is ``what is the correct state in the limit of 
$m_b\ra 0$ when there is a non-zero amount of particle-hole symmetric 
disorder?''
For that purpose let us consider the composite boson representation where 
the electrons are viewed as composite bosons carrying one quantum of 
fictitious magnetic flux each (i.e. the $\theta=1$ boson Chern-Simons theory). Here we recall that in this representation, 
the Bose superfluid phase corresponds to the $\nu=1$ quantum Hall liquid, and 
the Bose insulator (or the vortex superfluid) phase corresponds to the electron 
insulator. In between we can have a particular situation where the bosons and vortices are in the same state. The latter is marked by the so-called\cite{klz} ``self-duality condition'' where
\eq
(\rho_{xx}^b)^2+(\rho_{yx}^b)^2=\frac{1}{(\sxx^b)^2+(\sxy^b)^2}=(\frac{2\pi}{e^2})^2.
\label{dual}
\ee
To translate this condition into a statement concerning the electronic
response, we use
the connection formula between the electron and composite boson 
resistivity tensor,\cite{klz}
\eqa
&\rho_{xx}&=\rho_{xx}^b \nonumber \\
&\rho_{yx}&=\frac{2\pi}{e^2}+\rho_{yx}^b.
\label{bconnect}
\eea
With this identity, it is easy to see that
Eq.(\ref{dual}) is equivalent to Eq.(\ref{half})! 
Thus the particle-hole symmetric condition $\sxy=\frac{e^2}{4\pi}$ is equivalent to 
the statement of self-duality.\cite{shahar2} 

One example of the self-duality of the $\theta=1$ boson Chern-Simons theory is the critical point of the $\nu=0$ to $\nu=1$ plateau transition.\cite{klz} For the latter, it was argued that
\eqa
&&\sigma_{xy}^b=0\nonumber \\ 
&&\sigma_{xx}^b=\frac{e^2}{2\pi}.
\label{crit}
\eea 
We note that Eq.(\ref{crit}) constitutes a special solution to Eq.(\ref{dual}).
Values of the conductivity
consistent with Eq.(\ref{crit}) were found for both particle-hole symmetric and
non-symmetric disorder in numerical studies of non-interacting electrons
at this transition.\cite{bhatt}.
Recently, 
experiments have been performed which dramatically support the notion
that there is a universal resistivity tensor at the critical point, with
measured values in all cases
consistent with
the conjectured values of the composite boson conductivities (Eq.(\ref{crit})).\cite{note6,shahar}

The plateau transition (Eq.(\ref{crit})), being a critical point, obviously is infra-red unstable with respect to a single perturbation (which turns out to be $\sxy-\frac{e^2}{4\pi}$).
The fact that it is experimentally observable, implies that given the constraint that $\sigma_{xy}=e^2/4\pi$, it is
infrared stable. There are
infinitely many other possible solutions to Eq.(\ref{dual})\cite{note5}, with all of them consistent with $\sxy=\frac{e^2}{h}$; 
the
question is whether any of them corresponds to an
infrared stable fixed point 
in the presence of disorder (which, again, can be vanishingly small). 
If the answer is no, then even in the limit of vanishing disorder,
the ground-state of the system at $\nu=1/2$ is asymptotically equivalent
to the critical state at the $0\ra 1$ plateau transition. If the answer is yes, much new physics remains to be explored.  

\section{Perturbative results for composite fermion Hall conductivity in the absence of disorder:}

In this part of our paper we address the case where there is no disorder. 
As we stressed earlier, in that case the fact that 
$\sxy=\frac{e^2}{4\pi}$  does not uniquely determine the value of 
$\sxy^f$.  For example, so long as
$\sxx^f=\infty$, $\sxy^f$ can have any finite value. 
In particular, $\sxx^f=\infty$ 
and $\sxy^f=0$ is a perfectly legitimate solution.
 
In the following we shall compute $\sxy^f$ perturbatively.
The starting point of our subsequent analysis is 
the composite fermion Euclidean Lagrangian: 
\eqa
L[\psib,\psi,a]&=&\int d^2x\psib (\p_0+ieA_0-ia_0)\psi\nonumber \\
 -  &{1\over {2m_b}}& \int d^2x \psib({\vec{\nabla}}+ 
i\vec{A}-i\vec{a})^2\psi +
L_{a}[a],
\label{l}
\eea
where
\eqa
L_{a} & = & {1\over 8\pi^2\theta^2}\int d^2xd^2x'
[ b(x,t)-\bb ] v(x-x')[
b(x',t)-\bb ]\nonumber \\
&+&{i\over 4\pi\theta}
\int d^2x
\ep^{\mu\nu\lambda}a_{\mu}\p_{\nu}a_{\lambda}.
\label{lga}
\eea
$\psib$ and $\psi$ are the Grassmann fields associated with the composite 
fermions; $A_{\mu}$ and $a_{\mu}$ are the external and statistical 
gauge fields respectively;  
$b=\vec{\nabla}\times \vec{a}$; $v(x-x')$ is the bare 
interaction between electrons; $\bb\equiv 2\pi\theta\rb$ is the averaged 
statistical magnetic field. 
Moreover, we have made use of the Chern-Simons constraint that
$b(x,t)=2\pi\theta \rho(x,t)$.
By rescaling space, time, and the fermion fields, so that 
$x\ra k_F x, t\ra tk_F^2/m_b$, and $\psi,\psib\ra k_F^{-1}\psi, 
k_F^{-1}\psib$, (here 
$k_F\equiv\sqrt{\rb/\pi}$)
one can easily prove that in Eqs.(\ref{l}) and (\ref{lga}) the only 
dimensionless 
parameters are $\theta$, $\ap\equiv\hbar\w_c/E_c$, and $\rb \theta /2\pi B$, 
where $E_c=e^2/2\epsilon\sqrt{{\rb / \pi}}$, is the typical strength of the 
Coulomb interaction.  Here, we will consider only the case in which the 
magnetic field satisfies the
commensurability condition $2\pi\rb \theta/B=1$, so that
at the mean-field level the net effective magnetic field seen by 
the composite fermions is zero.

Within the class of models described by Eqs.(\ref{l}) and (\ref{lga}), 
the problem of physical interest corresponds to $\theta=2$, while 
the problem is simple in the limit $\theta \ra 0$ with $\ap$ fixed.
In that limit, the composite fermions are the bare electrons, 
and MFT is exact.\cite{stern} 
The question we are trying to address is ``what 
are the fluctuation corrections to this mean-field picture?''  
In carrying out the calculations, we choose
to work in Coulomb gauge, in which the gauge-field propagator, $D_{ij}$,
is a $2\times 2$ matrix, with $j=0, 1$ representing the time and space
components, respectively.

We have calculated $\s_{xy}^f$ perturbatively by evaluating the 
Feynman diagrams shown in Fig. \ref{sxy}. In 
that figure the wavy line represents the mixed gauge propagator $D_{01}$ 
and $D_{10}$. The open triangle, solid triangle, and square represent 
the density, 
current, and the diamagnetic vertices respectively. 
To lowest order in $\theta$ and $\ap$ we can used the bare gauge propagator,
$D_{00}(q_0,\vec{q})=V(\vec{q})$, $D_{11}=0$, and $D_{10}(q_0,\vec{q})=
D_{01}(q_0,\vec{q})=
i2\pi\theta/|q|$, where $V(q)$ is the Fourier transform
of $v(\vec{r})$. 
In this case, since $D_{01}$ does not depend on frequency, 
the integration can be easily done. Let $\w$ and $\vec{q}$ be the external 
frequency and momentum respectively.  We have looked at two limits: i) 
$|\vec{q}|\ra 0$ first and $\w\ra 0$ second (this is the canonical limit 
for defining the conductivities), and ii) $\w\ra 0$ first and $|\vec{q}|\ra 0$
second. In case i) all the individual graphs shown in Figs.1(a)-1(d) are zero. 
In case ii) the contributions to $\s_{xy}^f$ from Figs.1(a), 1(b), 
and 1(c),1(d) are $\pm\theta e^2/16\pi$ respectively; thus the net result is 
again zero. (We note that for this case we found that the characteristic 
momentum carried by the gauge line is of order $k_{F}$.)
Since $|\vec{q}|\ne 0$ breaks Galilean invariance,
we regard the value of $\sxy^f$ in 
case ii) as a more stringent test of whether the time reversal symmetry of composite fermions is restored.

Therefore to this order we obtain $\sxy^f=0$. This result is  {\it consistent} 
with the notion of a composite Fermi liquid in zero magnetic field. {\it If} 
this were true to all orders, i.e. $\sxy^f=0$ in the absence of disorder, 
we would be left with the following
situation: In the limit of $m_b\ra 0$, the composite fermion Hall conductance 
in the presence of 
particle-hole symmetric disorder of vanishing strength 
would differ by $-e^2/4\pi$ from its value in
the absence of disorder. 
Singular behavior in the limit of zero disorder is, of course, not unprecedented;
For non-interacting electrons in two dimensions, the zero-temperature
conductivity is infinite in the absence of disorder, and 0 (due to
localization) for arbitrarily small disorder. 
However, this behavior is due to a subtle, infra-red instability of the
Fermi-liquid fixed point in two dimensions, and can be circumvented by
considering a finite size system, or a system at finite temperature, in
which case the zero disorder and vanishing disorder results coincide.
The situation for the composite fermions is, we believe, fundamentally
different. This is because the fact that particle-hole symmetry implies 
$\sxy=\frac{e^2}{4\pi}$ does not rely on either
the zero temperature or the thermodynamic limits;
the fact that $\sxy^f\ne 0$ is not a delicate infrared phenomenon. It 
merely reflects the fact that through the flux-density correlation imposed by 
the Chern-Simons term, disorder makes flux cancellation an impossible task. 
For this reason, {\it we believe} (without proof) that the results in the
presence of weak disorder are pertinent to understanding the properties of
the system in the absence of disorder.

\section{Digression: other perturbative results}

In this section we report some other perturbative results we have obtained. 
These results do not directly address 
the question of $\sxy^f$ in the pure system, but do shed some light on
other properties of this system.

We shall concentrate on the density-density and 
current-current correlation functions. 
The effects of the bare $D_{00}$ are 
identical to those of a static Coulomb interaction. 
As is customary in this case, a RPA resumation is
performed to 
screen $D_{00}$ and $D_{11}$. If one uses the renormalized $D_{00}$ 
and $D_{11}$ to compute the 1-loop corrections to the composite fermion 
self-energy, $\Sigma(q_0,\vec{q})$, 
the contribution from longitudinal fluctuations, {\it i.e.} those which 
involve $D_{00}$, diverges logaritmically with the size of the system for 
fixed $q_0$ and $\vec{q}$.\cite{hlr}
The contribution from transverse fluctuations, {\it i.e.}
those involving  $D_{11}$, are regular in the system size, but contribute a 
logarithmicly diverging correction to the 
effective mass.\cite{hlr} 

In the following we shall prove that to the same level of 
approximation in  the density-density and current-current correlation 
functions (see Fig.\ref{rr}), the {\it divergent}  contribution
to the self-energy from longitudinal gauge field fluctuations
is exactly canceled by the corresponding vertex correction for 
{\it all} $\vec{q}$ and $\w$.  However, a logarithmic singularity 
from transverse gauge fluctuations stays in these response functions,
but {\it only} at wave-vector $q=2k_F$.\cite{ioffe,stamp}

To begin with, let us recall the origin of the divergent contribution to
the composite fermion self-energy.\cite{hlr} 
Since the divergence originates from the high-energy, small momentum region 
($\omega>>q$), we can use the following expression for the
RPA screened $D_{00}$:
\begin{equation}
D_{00}(q)=\frac{2 \pi\theta}{q^2} \frac{q^2_0 \omega_c}{q^2_0+ \omega_c^2}. 
\end{equation}
Substituting this expression into the formula for the lowest  order self-energy 
correction, we obtain:
\begin{equation}
\Sigma(k) = - \int \frac{d^3 q}{(2\pi)^3} G_0(k-q) D_{00}(q).
\label{se}
\end{equation}
>From Eq.(\ref{se}) a singular term can be extracted:
\begin{equation}
\Sigma(k)= \frac{\omega_c \theta}{2} ln (q_{max}L) \frac{\varepsilon_k - ik_0}{\omega_c + sgn(\varepsilon_k)(\varepsilon_k -k_0)}.
\end{equation}

Now let us consider the analogous contribution to the 
density-density correlation 
function, $<\rho \rho>$.  To the lowest order $<\rho \rho>$ is given by 
the sum of three diagrams  Fig.\ref{rr}(a, b, c). The corresponding 
analytical expressions are:
\begin{eqnarray}
\Pi_1 &=& - \int \frac{d^3 p}{(2\pi)^3} G_0(p) G_0^2(p+q) 
\Sigma(p+q), \label{pi1} \\
\Pi_2 &=& - \int \frac{d^3 p}{(2\pi)^3} G_0^2(p) G_0(p+q) 
\Sigma(p), \label{pi2} \\
\Pi_3 &=& - \int \frac{d^3 p}{(2\pi)^3} G_0(p) G_0(p+q) 
\Gamma(p,p+q), \label{pi3},
\end{eqnarray}
where the dressed vertex $\Gamma$ is given by
\begin{equation}
\Gamma(p,p+q) = - \int \frac{d^3 k}{(2\pi)^3} G_0(p+q+k) G_0(p+k) D_{00}(k).
\label{g}
\end{equation}
It is straightforward to show that the following relation holds between the divergent contributions to $\Sigma$ and $\Gamma$:
\begin{equation}
\Gamma(p,p+q) = \frac{\Sigma(p) - \Sigma(p+q)}{iq_0 - \varepsilon_{p+q} + \varepsilon_{p}}.
\end{equation}
Substituting this expression into Eqs.(\ref{pi1}-\ref{pi3}), and using the identity
\begin{equation}
G_0(p) G_0(p+q)= \frac{G_0(p) - G_0(p+q)}{iq_0 - \varepsilon_{p+q} + \varepsilon_{p}},
\end{equation}
one can show that:
\begin{equation}
\Pi =\sum_{i=1}^3\Pi_i= - \int \frac{d^3 p}{(2\pi)^3} \frac{ G_0^2(p) \Sigma(p) - G_0^2(p+q) \Sigma(p+q)}{iq_0 - \varepsilon_{p+q} + \varepsilon_{p}}.
\end{equation}
The above expression vanishes after integration 
over $p_0$ due to the analytical structure of the integrand, that is because the 
poles of $G_0(p)$ and $\Sigma(p)$ are on the same side of the real axis.

Now we turn to the current-current correlation function 
$\Pi^{\ap\bt}=<j_{\alpha}j_{\beta}>$. 
To get $\Pi_1^ {\alpha\beta}, \Pi_2^{\alpha\beta}, \Pi_3^{\alpha\beta}$ 
associated with the diagrams in Fig.\ref{rr}(a,b,c), we need to insert 
current vertices $p_{\alpha}(p+q)_{\beta}$, $p_{\alpha}(p+q)_{\beta}$ into 
Eqs.(\ref{pi1},\ref{pi2}), and $(p+k)_{\alpha}(p+q)_{\beta}= 
p_{\alpha}(p+q)_{\beta} + k_{\alpha}(p+q)_{\beta}$ into Eq.(\ref{g}). 
Since the $k_{\alpha}(p+q)_{\beta}$ term in the last expression does not 
produce any divergence, it can be neglected. Therefore  to obtain the singular 
contributions to $\Pi_1^ {\alpha\beta}, \Pi_2^{\alpha\beta}, 
\Pi_3^{\alpha\beta}$, all we need to do is to multiply the integrands 
in Eq.(\ref{pi1}-\ref{pi3}) by the factor $p_{\alpha}(p+q)_{\beta}$. 
Since the last operation does not affect the pole structure; the proof 
goes through as before and the singular contribution again vanishes.

Now, we turn to the contributions to these correlation functions from
the transverse gauge-field fluctuations, where 
the singular 
contributions from the self-energy and vertex corrections do not cancel 
for $\mid\vec{q}\mid=2k_F$. (They do cancel at other 
$\mid\vec{q}\mid$.\cite{ioffe,stamp,kim}) The graphs used in that 
calculation are summarized in Fig.\ref{rr}. The result for the $2k_F$ 
density-density correlation function is given by
\eq
\Delta\Pi (\omega, 2k_F)/
\Delta\Pi_0 (\omega,2k_F)
=1+\theta\alpha C_1 ln\left[{E_F\over {\mid \omega\mid}}\right]. 
\label{eq:DeltaPi}
\ee
Here $\Pi(\w,q)\equiv\int d^2x dt e^{i(\w t-\vec{q}\cdot\vec{x})}
<T[\rho(x,t)\rho(0,0)]>$ with 
$\Pi_0$ being the density-density correlation function of free electrons,
$\Delta\Pi(\w,q)=\Pi(\w,q)-\Pi(0,q)$, 
and $C_1 =-{1\over {\pi}}[{1\over 4}+ln({{\alpha\theta}\over 2})]$.
The above result shows that small $\ap$, hence {\it strong} Landau level 
mixing, tends to stabilize the composite fermion mean-field theory 
against divergent corrections arising from the transverse gauge
field fluctuations. 

To summarize, the perturbative results for the pure system are 
consistent with the existence of time reversal symmetry in the long
wave-length low energy properties of the composite fermions. 
The only sign of non-Fermi liquid behavior is Eq.(\ref{eq:DeltaPi}). 
Whether this singularity signals a true asymptotic state that lacks 
time reversal symmetry, cannot be determined on the basis of our results. 
However, along these lines,
we would like to point out a {\it possible} implication of 
Eq.(\ref{eq:DeltaPi}), i.e. that
at finite Landau level mixing, there exists a crossover temperature scale, 
$T_{cr}$, above which all divergent corrections to composite
Fermi liquid behavior are numerically insignificant.\cite{ioffe} 
If this interpretation is correct, then Eq.(\ref{eq:DeltaPi})
suggests that this temperature is exponentially small in the limit of a 
large amount of Landau level mixing, so there would exist a broad
temperature range, $T_{cr} \ll T \ll \omega_c$, in which Fermi liquid
behavior would be observable.

Finally, the fact that the divergent self-energy correction from $D_{00}$ is 
canceled by the vertex correction for {\it all} external momenta leads us
to the following tentative conclusion:
single composite 
fermion excitations ( and by extenuation, probably any excitation with
net ``statistical charge'') are {\it not} part of the physical spectrum. 
Instead, the physical excitations are {\it statistical 
charge-neutral} particle-hole 
excitations.\cite{kim}
Of course, we have proven the consistency of this viewpoint only 
to lowest order in perturbation theory, so at this point we can only conjecture that it remains valid more generally.

\section{Final discussions:}
In this section we restore $e,c,\hbar$ and $k_B$. 
For real systems there {\it is} appreciable Landau level mixing ($\ap\sim 1$). 
Thus an important question is ``what does the LLL and particle-hole symmetry 
constraint having to do with reality?'' (Here we should stress that although 
for real systems, it is not clear that $\sxy^f=-{1\over 2}\frac{e^2}{h}$, but
the general observation that disorder destroys flux cancellation and hence 
makes $\sxy^f\ne 0$ should still be generically true.) 

One way to address this question, is to examine it in the light of some
recent experimental results of Wong and Jiang 
\cite{hwj}. In that
study\cite{hwj} Wong and Jiang have attempted to map out the nature of the
global, zero temperature phase diagram in 
the density-filling factor plane in the neighborhood of $\nu=1/2$ using 
gated GaAs heterojunctions with mobilities in the range 
$\mu \le 2\times 10^6 cm^2/Vs$.  (Since the mobility is a monotone increasing
function, it is useful to think of varying the density as varying the
degree of disorder.)
Wong and Jiang have identified a line, which can be unambiguously 
associated with
the $0\ra 1$ plateau transition, on which the full conductivity tensor
(or resistivity tensor) is apparently temperature independent;  moreover,
everywhere on this line, $\sigma_{xx}\approx
\sigma_{xy} \approx (1/2) e^2/h$, consistent with theoretical expectations.
This line lies at $\nu\approx 1$ in the low mobility (high disorder) limit,
and approaches $\nu=1/2$ as the disorder is decreased.
In the highest mobility samples, however, the boundary of the $\nu=1$ phase
can no longer be clearly identified,
possibly
due to finite temperature effects.  (The lowest temperature
in this experiment is  50mK.)  A similar line has been identified
at $\nu < 1/2$
corresponding to the $0\ra 1/3$ plateau transition, on which
$\sigma_{xx}\approx (1/10) e^2/h$ and $\sigma_{xy}\approx(3/10) e^2/h $, 
independent of temperature and density.  In addition to these familiar
phase-boundaries, two other characteristic behaviors have been observed,
which can be used to map out lines in the phase diagram of, as yet,
undetermined meaning.
One such line is more-or less parallel to the density axis at 
$\nu \approx 1/2$, and occurs only at relatively high mobilities:  
On this line, $\sxy=(1/2)e^2/h$,
independent of temperature and density, while 
$\sigma_{xx}$ varies with density, and is still temperature dependent,
even at the lowest temperatures. At the low density end of this line,
$\sigma_{xx}$ approaches $(1/2)h/e^2$ and becomes ever more
weakly temperature dependent, {\it i.e.} this line is apparently
the continuation
of the
$0\ra 1$ phase boundary. 
However, for high density samples, the magnitude of $\sigma_{xx}$ is 
about $(0.08)e^2/h$ at the lowest temperatures, where it is still quite
noticeably temperature dependent.
This result suggests that $\sxy=e^2/2h$, i.e. particle-hole symmetry, 
is more robust than the universal dissipative transport. (This is also
consistent with recent experiments on the non-linear transport near
quantum Hall transitions\cite{shahar2} which reveal that a form of
self duality (which for the $1\ra 0$ transition is the same as
particle-hole symmetry) is observed over a much wider range of
filling factors than the critical behavior, itself.)  It remains to be 
seen whether upon further cooling $\sigma_{xx}$ rises to the universal value
(as would be expected if this is indeed the continuation of the $0\ra1$
phase boundary).
Finally, another line is observed on which
$\rho_{xy}\approx 2h/e^2$, and is approximately temperature independent. 
(These two lines necessarily converge in the high mobility limit,
as $\rho_{xx}\ra 0$.)  Along this line $\sxy^f=0$.

In a recent preprint\cite{ssh} Simon, Stern and Halperin have
pointed out a difficulty 
in the mean-field, and what they call the $(M)RPA$, approximations for the 
composite Fermi liquid theory.
They consider the limit of $m_b\ra 0$, and an inhomogeneous external 
magnetic field $B(r)=B_{1/2}+\delta B(r)$. From the electron point of view, 
due to the zero point kinetic energy $\hbar\w_c(r)$, the region of $\delta 
B(r)<0$ will be populated by electrons while that of $\delta B(r)>0$ will 
not. From the composite fermion point of view the same physics {\it is} 
reflected in energy associated with the zero-point composite fermion density fluctuation. Instead, in the spirit of Landau theory,
Simon, Stern, and Halperin suggested modifying the composite Fermi liquid
theory by attaching a magnetic moment, of determined strength, to each
composite Fermion.
After taking into account  the 
magnetization current associated with this moment, they arrived at a new 
approximation - the $M^2RPA$. While it seems to us unlikely that this same correction
will simultaneously correct the value of $\sxy^f$ in the presence of disorder\cite{simon}, it is possible that a similar in spirit modification of
the basic constituents of the composite Fermi liquid theory might exist that would accomplish this task.
%
\vspace{0.2in}
\\
{\bf Acknowledgments:}  We acknowledge illuminating
discussions with E.~Fradkin, L.~Ioffe, A.~C.~Neto, P. Stamp, 
and X-G.~Wen.  We thank Dr. D. Khveshchenko for pointing out an error in our earlier perturbative result for $\s_{xy}^f$. SK was
supported in part by the NSF under grant number DMR93-12606 at UCLA, and
by a Miller Fellowship at UCB.

\begin{center}
\bf Figure Captions
\end{center}
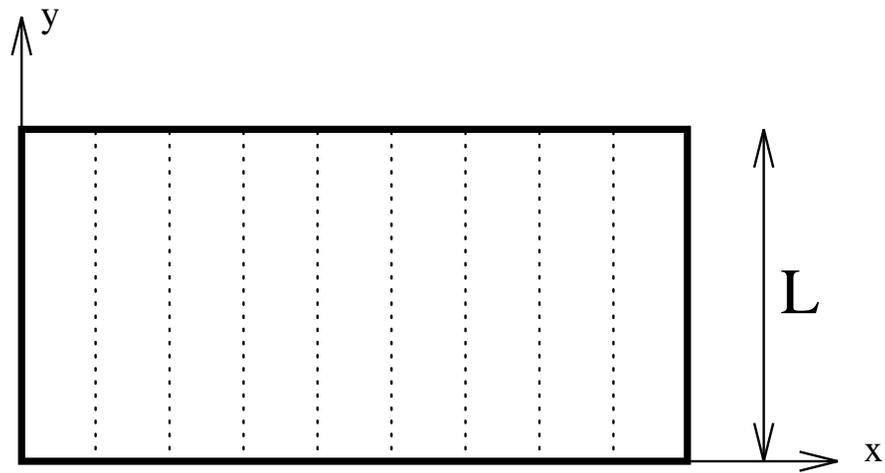
\begin{figure}
\caption{The geometry for the quantum Hall system.} 
\label{hallbar}
\end{figure}

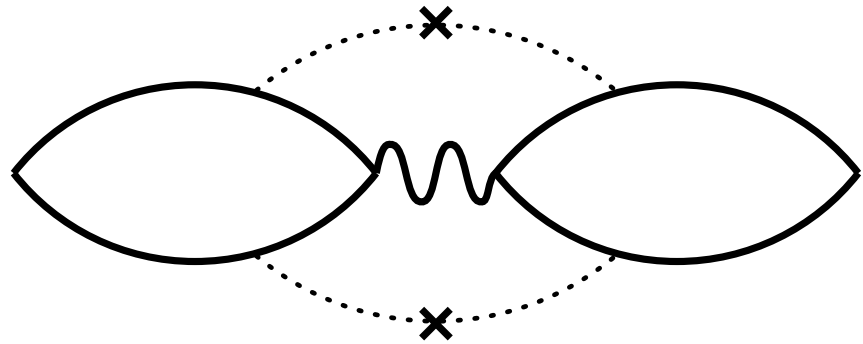
\begin{figure}
\caption{An example where an impurity averaged irreducible bubble diagram does not appear after averaging the irreducible bubble diagrams for specific disorders.} 
\label{irrbubble}
\end{figure}

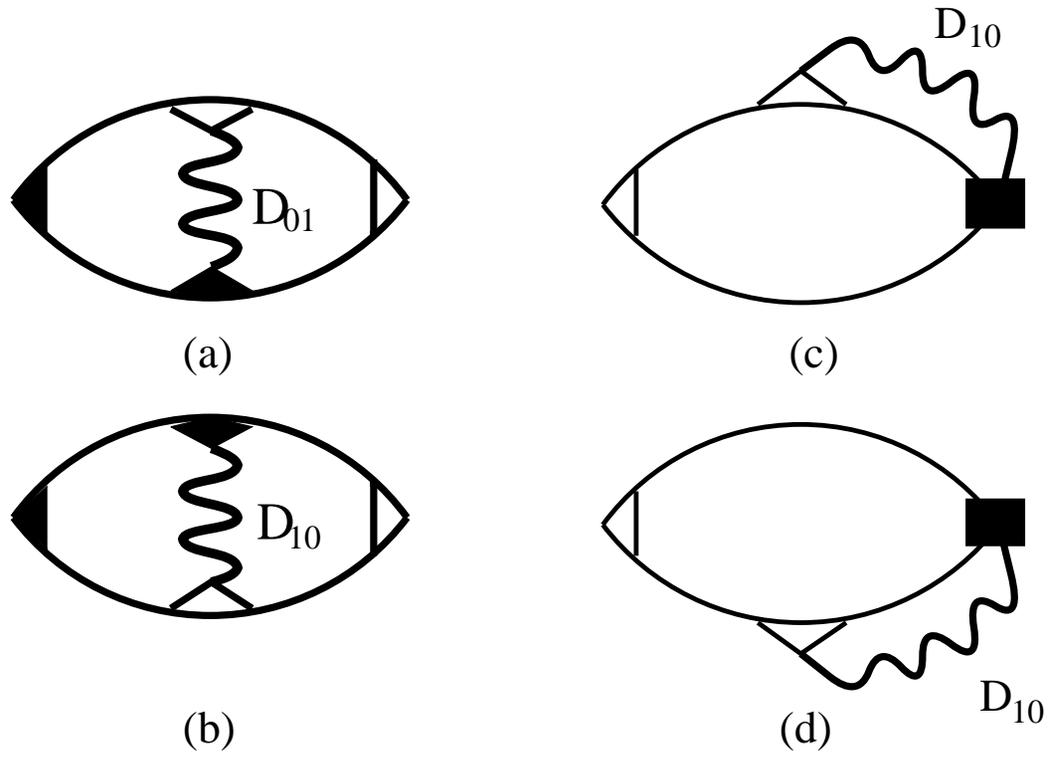
\begin{figure}
\caption{Feynman diagrams for $\s_{xy}^f$. For $q\ra 0, \w\ra 0$ while $\w>>q$ each individual graphs in a)-d) vanishes.
For $q\ra 0, \w\ra 0$ while $\w<<q$ the diagrams a),b) cancels diagram c), d). The diagrams corresponding to self-energy insertions vanish due to symmetry.} 
\label{sxy}
\end{figure}

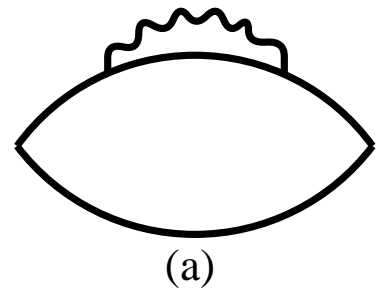
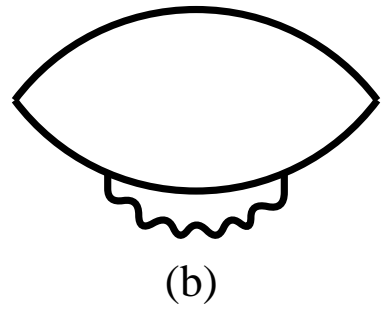
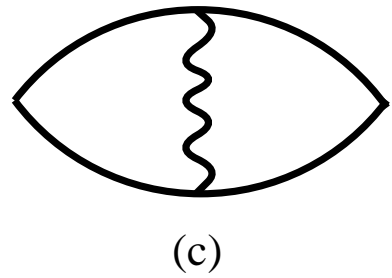
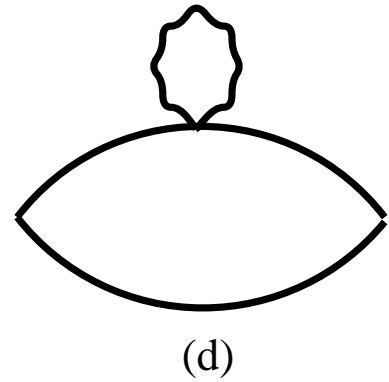
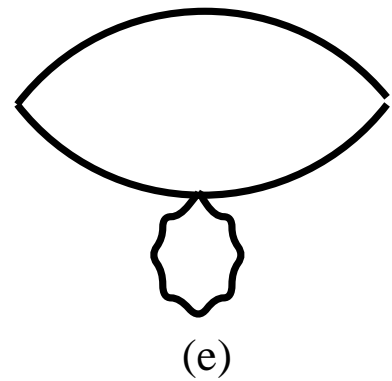
\begin{figure}
\caption{Feynman diagrams for $\Pi(q_0,\vec{q})$. For logitudinal gauge fluctuations, diagrams d) and e) are absent.}
\label{rr}
\end{figure}

\vspace*{\fill}

\begin{references}
\bibitem{jiang} H.W.Jiang et al., Phys.Rev.Lett. {\bf65}, 633 (1990). 
\bibitem{willet} R.L.Willet et al.,Phys.Rev.Lett. {\bf71}, 3846 (1993). 
\bibitem{hlr}  B. I. Halperin, P. A. Lee, and N. Read, Phys. Rev. B {\bf47}, 7312 (1993).
\bibitem{kz} V.Kalmeyer and S.C.Zhang, Phys.Rev.B {\bf46}, 9889 (1992). 
\bibitem{jain} J.K.Jain, Phys.Rev.Lett. {\bf63}, 199 (1989).
\bibitem{zhk} a) S.C. Zhang, H. Hanson and S. Kivelson, Phys. Rev. Lett. 
{\bf62}, 82 (1989); {\bf62}, 980 (E) (1989). b) A.~Lopez and E.~Fradkin,
Phys. Rev. B {\bf 44}, 5246 (1991). 
\bibitem{stern} A. Stern and B.I. Halperin, Phys. Rev. B {\bf52}, 5890 (1995).
\bibitem{wilczek} C. Nayak and F. Wilczek, Nucl.Phys.B {\bf417}, 359 (1994).
\bibitem{marsden} H-J Kwon, A. Houghton and J.B. Marston, Phys. Rev. B {\bf52}, 8002 (1995).
\bibitem{ioffe} B.L. Altshuler, L.B. Ioffe and A.J. Millis, Phys. Rev. B {\bf50}, 14048 (1994).
\bibitem{stamp} D.V. Khveshchenko and P.C.E.Stamp, Phys.Rev.Lett. {\bf71}, 2118 (1993); Phys.Rev.B {\bf49}, 5227 (1994).
\bibitem{fsurf} W.Kang et al., Phys.Rev.Lett. {\bf71}, 3846 (1993);
V.J. Goldman, B.Su and J.K. Jain, Phys. Rev. Lett. {\bf72}, 2065 (1994); 
R.L.Willet, K.W.West and L.N.Pfeiffer, Phys.Rev.Lett. {\bf75}, 2988 (1995).
\bibitem{note} In the presence of disorder potential $U(x)$, 
we speak of p-h symmetry if the disorder ensemble has the property that 
$P[U(x)]=P[-U(x)]$, where $P[U]$ is the probability that a particular 
$U(x)$ is realized. Here,
without loss of generality, we have set $\int d^2x U(x)=0$.
\bibitem{girvin} S.M. Girvin, Phys. Rev. B {\bf29}, 6012 (1984).
\bibitem{sk}  S.~L.~Sondhi and S.~A.~Kivelson, Phys. Rev. B {\bf46}, 13319 (1992).
\bibitem{sd} R.Rajaraman and S.L.Sondhi, Int. J. Mod. Phys. B {\bf8}, 1065 (1994).
\bibitem{skk}  S.~Sondhi, A.~Karlhede, S.~A.~Kivelson, and E.~H.~Rezayi, Phys. Rev. B {\bf47}, 16419 (1993).
\bibitem{note2} Here the integration over space is essential in order to show Eqs.(\ref{jpx})and (\ref{jpy}) given Eq.(\ref{jp}).
\bibitem{klz} D.-H. Lee, S.A. Kivelson and S.-C. Zhang, Phys. Rev. Lett. {\bf67}, 3302 (1991); S.A. Kivelson, D.-H. Lee and S.-C. Zhang, Phys.Rev.B {\bf46}, 2223 (1992).
\bibitem{shahar2}  D. Shahar {\it et al}, Cond-Mat/9510113.
\bibitem{bhatt} Y. Huo, R.E. Hetzel and R.N. Bhatt, Phys. Rev. Lett. {\bf70}, 481 (1993).
\bibitem{note6} To address other integer or fractional plateau transitions, the $\theta=1$ Chern-Simons theory has to be modified appropriatly. In Ref.\cite{klz} it was proposed that all of these transitions are marked by Eq.(\ref{crit}), where $\sxx^b$ and $\sxy^b$ are the conductivities of the ``order parameter'' composite boson.
\bibitem{shahar}D. Shahar, D.~C. Tsui, M. Shayegan, R.~N. Bhatt, and
 J.~E. Cunningham, Phys. Rev. Lett. {\bf74}, 4511 (1995).
\bibitem{note5} One noteable example is $\sxx^b=0$ and $\sxy^b=\frac{e^2}{2\pi}$. Upon using Eq.(\ref{bconnect}) it  
it becomes $\r_{xx}=0$ and $\r_{yx}=\frac{4\pi}{e^2}$, i.e. the resistivity tensor of the pure system.
By using the connection formula between composite boson and 
composite fermion resistivities
$\r_{xx}^b=\r_{xx}^f$, and $\r_{yx}^b=\r_{yx}^f+\frac{2\pi}{e^2}$, one can show that 
all other solutions require $\sxy^f=-{e^2}/{4\pi}$! 
\bibitem{kim} Y.B. Kim, A. Furusaki, X-G Wen and P.A. Lee, Phys. Rev. B {\bf50}, 17917 (1994).
\bibitem{hwj}L.~W.Wong and H.W. Jiang, unpublished.
\bibitem{ssh} S.H. Simon, A. Stern and B.I. Halperin, preprint.
\bibitem{simon}  S.~Simon and Y.B. Kim, private communication.

\end{references}
\end{document}